\documentclass[twocolumn,aps,superscriptaddress,preprintnumbers]{revtex4}

\usepackage{comment}
\usepackage{latexsym}
\usepackage{hyperref}
\usepackage{amsmath,amssymb,amsfonts}
\usepackage[caption=false]{subfig}
\usepackage{bm}
\usepackage{graphics,xcolor}
\usepackage{epstopdf}
\usepackage{multirow}
\usepackage{psfrag}
\usepackage[latin1,utf8]{inputenc}
\usepackage{verbatim}
\usepackage{tikz}
\usepackage{braket}
\usepackage{bm}
\usepackage{longtable}
\usepackage[export]{adjustbox}
\usepackage{enumerate}
\usepackage{hyperref}

\usepackage{tabularx}
\usepackage{xcolor}
\newcommand{\notablue}[1]{{\color{blue}\textrm{#1}}}

\graphicspath{{figuras/}}
\DeclareUnicodeCharacter{2212}{-}
\DeclareGraphicsExtensions{.pdf,.png,.jpg,.gif,.eps}
\hypersetup{
    colorlinks=true,       
    linkcolor=blue,        
    citecolor=blue,        
    filecolor=blue,        
    urlcolor=blue,         
    runcolor=blue
}
\newcommand{\matelement}[3]{\langle #1 | #2 | #3 \rangle}

\newcommand{\threejm}[6]{ \left(\begin{array}{ccc} #1 & #3 & #5\\
                                              #2 & #4 & #6
                                \end{array}
                          \right)}
\newcommand{\sixj}[6]{ \left\{\begin{array}{ccc} #1 & #3 & #5\\
                                              #2 & #4 & #6
                                \end{array}
                          \right\}}

\def\xi{\mathbf{x}_i}

\begin{document}

\title{$C_6$ coefficients for interacting Rydberg atoms and alkali-metal dimers}

\author{Vanessa Olaya}
\affiliation{Department of Physics, Universidad de Santiago de Chile, Av. Ecuador 3493, Santiago, Chile.}

\author{Jes\'{u}s P\'{e}rez-R\'{i}os}
\affiliation{Fritz-Haber-Institut der Max-Planck-Gesellschaft, Faradayweg 4-6, 14195 Berlin, Germany}

\author{Felipe Herrera}
\email{felipe.herrera.u@usach.cl}
\affiliation{Department of Physics, Universidad de Santiago de Chile, Av. Ecuador 3493, Santiago, Chile.}
\affiliation{Millennium Institute for Research in Optics MIRO, Chile.}

\date{\today}


\begin{abstract}
We study the van der Waals interaction  between Rydberg alkali-metal atoms with fine structure ($n^2L_j$; $L\leq 2$) and heteronuclear alkali-metal dimers in the ground rovibrational state ($X^1\Sigma^+$;  $v=0$, $J=0$).  We compute the associated $C_6$ dispersion coefficients of  atom-molecule pairs involving $^{133}$Cs and $^{85}$Rb  atoms interacting with KRb, LiCs, LiRb, and RbCs molecules. The obtained dispersion coefficients can be accurately fitted to a state-dependent polynomial $O(n^7)$ over the range of principal quantum numbers $40\leq n\leq 150$. For all atom-molecule pairs considered,  Rydberg states $n^2S_j$ and $n^2P_j$ result in attractive $1/R^6$ potentials. In contrast, $n^2D_j$ states can give rise to repulsive potentials for specific atom-molecule pairs. The  interaction energy at the LeRoy distance approximately scales as $n^{-5}$ for $n>40$. For intermediate values of $n\lesssim40$, both repulsive and attractive interaction energies in the order of $ 10-100 \,\mu$K can be achieved with specific atomic and molecular species.  The accuracy of the reported $C_6$ coefficients is limited by the quality of the atomic quantum defects, with relative errors $\Delta C_6/C_6$ estimated to be no greater than 1\% on average. 
\end{abstract}

\maketitle


Rydberg atoms with high principal quantum number have exaggerated properties such as orbital sizes of thousands of Bohr radii, long radiative lifetime exceeding microseconds and are extremely sensitive to static electric fields. The latter is due to their giant dipole moment and electric polarizability. These exotic features have made Rydberg atoms a widely studied platform for fundamental studies and applications such as quantum information processing \cite{Saffman2010}, quantum nonlinear optics \cite{Peyronel2012}, and precision measurements \cite{Facon2016}. 
Ultracold Rydberg atoms can be trapped with high densities \cite{Liebisch}, enabling studies of long-range interactions \cite{Kamenski,Mourachko} and resonant excitation exchange (F\"oster) processes \cite{Anderson}. F\"oster energy transfer has been proposed as a non-destructive method for detecting cold molecules in optical traps \cite{Jarisch,Zeppenfeld} that can  simultaneously trap Rydberg atoms and molecules. These hybrid Rydberg-molecule systems have been considered for direct sympathetic cooling of cold molecules into the ultracold regime through long-range collisions \cite{Zhao,Huber}. There is also a growing interest in the preparation and study of ultralong-range Rydberg molecules \cite{Perez2015,Greene1,Bendkowsky,Shaffer2018}, which emerges from the binding of a Rydberg alkali-metal atom and a neutral molecule in the long range. Rydberg molecules have extraordinary properties that can be controlled by driving the Rydberg electron and may lead to promising applications \cite{Eiles,Luukko}.

 For an atom-molecule system at low kinetic energies, the relevant scattering properties at distances beyond the LeRoy radius \cite{LeRoy}, are determined by the long-range interaction between particles \cite{Perez2015}. For $R$ being the relative distance of the centers of mass of the colliding partners, the long-range interaction potential can be written as an expansion of the form $V(R) = \sum_n C_n/R^{n}$, with $n\geq 3$ for neutral particles. For collisions between particles in their ground states, the lowest non-vanishing van der Waals coefficient is $C_6$. For neutral particles in excited entrance channels, the coefficients $C_3$ and $C_5$ can also contribute to the potential. 

In this work, we report a large set of van der Waals $C_6$ coefficients that determine the long-range interaction between selected  heteronuclear alkali-metal dimers (LiCs, RbCs, LiRb and KRb) in their electronic and rovibrational ground state ($^1\Sigma^+, v=0, J=0$) with $^{85}$Rb and $^{133}$Cs atoms in Rydberg states $n^2L_j$ with $15 \leq n \leq 150$, taking into account spin-orbit coupling.  $n$ is the atomic principal quantum number, $L$ is the atomic orbital angular momentum, $j$ is the total electronic angular momentum, $\nu$ the vibrational quantum number of the molecule and $J$ is the rotational angular momentum. For every  atomic state considered, $C_5$ vanishes exactly by dipole selection rules and resonant $C_3$ contributions can be ignored. The precision of the computed $C_6$ coefficients is  limited by the accuracy of the semi-empirical quantum defects used. 

We fit the computed atom-molecule $C_6$ coefficients as a function of $n$ to a polynomial of the form
\begin{equation}\label{eq:fitting}
C_6= \gamma_0+ \gamma_{4} \,  n^4 + \gamma_{6} \, n^6 + \gamma_{7} \, n^7, 
\end{equation}
which is valid in the range $40\leq n\leq 150$. The accuracy of the fitting increases with $n$. We list the fitting coefficients for all the atom-molecule pairs considered in Table \ref{tab:Cs fit} for Cesium and \ref{tab:Rb fit} for Rubidium. The proposed $n^7$ scaling is consistent with the scaling of the static Rydberg state polarizability \cite{Gallagher}. The data set of computed $C_6$ coefficients is provided in the Supplementary Material. 

We describe in Sec. \ref{Method} the theoretical and numerical methodology used to compute $C_6$ coefficients. In Sec. \ref{Results} we present the dispersion coefficients for selected atom-molecule pairs, and discuss their accuracy in Sec. \ref{discussion}. We conclude by discussing possible implications of our results. 

\section{Methodology}\label{Method}

In this section, we briefly review the theory of long-range interaction between a heteronuclear alkali-metal dimer (particle $A$) and an alkali-metal atom (particle $B$) in an arbitrary fine structure level $n^2L_j$, in the absence of external static or electromagnetic fields. Our work extends the results in Refs. \cite{Lepers3,Lepers2,Lepers} to fine structure states with high $n$, as relevant for Rydberg states.

\subsection{Interaction potential}\label{subsec:potential}

Consider the charge distributions of molecule $A$ and atom $B$, separated by a distance larger than their corresponding LeRoy radii \cite{LeRoy}. The long-range electrostatic interaction between a molecule ($A$) and atom ($B$) is given by the multipole expansion \cite{Flannery}

\begin{equation}\label{eq:multipole expansion}
\hat{V}_{AB}(R) = \sum_{L_A=0}^{\infty}\sum_{L_B=0}^{\infty}\sum_{q=-L_{<}}^{L_{<}}\frac{f_{L_AL_Bq}}{R^{1+L_A+L_B}}\hat{Q}^{L_A}_q(\hat r_A)\hat{Q}^{L_B}_{-q}(\hat r_B),
\end{equation}
where $L_<$ is the smallest of the integers $L_A$ and $L_B$. The multipole moments $\hat{Q}^{L_{X}}_q(\hat r_{X})$ associated with a particle $X = (A, B)$ can be written  in spherical tensor form \cite{Zare}

\begin{equation}
\hat{Q}^{L_X}_q(\hat r_{X}) = \bigg( \frac{4\pi}{2L_X+1} \bigg)^{1/2}\sum_{i} q_i \hat{r}_i^{L_{X}} Y_{q}^{L_{X}}(\theta_i,\varphi_i),
\end{equation}
where $q_i$ is the $i$-th charge composing the $X$-distribution and $Y_q^{L_X} (\theta_i,\varphi_i)$ is a spherical harmonics. Expectation values of the multipole moments depend on the electronic structure of the particle. For a two-particle coordinate system in which the quantization axis  is pointing from the center of mass of particle $A$ to the center of mass of particle $B$, the factor $f_{L_AL_Bq}$ in equation (\ref{eq:multipole expansion}) becomes \cite{Lepers}

\begin{equation}
 f_{L_A L_B q}=\frac{(-1)^{L_B}(L_A+L_B)!}{\sqrt{(L_A+q)!(L_A-q)!(L_B+q)!(L_B-q)!}}.
\end{equation}

\subsection{Dispersion coefficients}

In the long range, the interaction Hamiltonian $\hat V_{AB}(R) $ in Eq. (\ref{eq:multipole expansion}) gives a perturbative correction to the asymptotic energies of the collision partners. This energy shift is the interaction potential $V_{AB}(R)$, which can be evaluated using  second-order  degenerate perturbation theory  to read \cite{Marinescu,Marinescu1}
\begin{equation}\label{eq:dispersion potential}
V_{AB}(R) = \sum_n \frac{C_n}{R^n},
\end{equation}
where $C_n$ are the dispersion coefficients.  Values of $C_n$ are obtained by defining the zero-th order eigenstates of the collision pair. In the absence of external fields, these are given by product states of the form $\ket{\Phi_{AB}^{(0)}}=\ket{\Psi_A^{(0)}}\ket{\Psi_B^{(0)}}$, where in our case $\ket{\Psi_A^{(0)}} \equiv \ket{X^1\Sigma^+}\ket{v=0,J=0}$ is  the absolute ground state of an alkali-metal dimer and $\ket{\Psi_B^{(0)}} \equiv \ket{(n^2L) jm}$ is a general fine structure state of an alkali-metal atom, with $m$ being the projection of the total electronic angular momentum along the quantization axis. 

The non-degenerate rovibrational ground state of a $^1\Sigma$ molecule has a definite rotational angular momentum, thus parity. Therefore, the lowest non-zero contribution to the expansion in Eq. (\ref{eq:dispersion potential}) is $C_6/R^6$. For molecules in an excited rotational state $J\geq 1$, quadrupole moments can give non-vanishing $C_5$ coefficients \cite{Lepers3,Lepers2,Lepers}. In this work, we only consider the ground rotational state (i.e., $C_5=0$).  The second-order atom-molecule dipole-dipole interaction thus leads to a $C_6$ coefficient of the form 
\begin{eqnarray}\label{eq:2do-order}
C_6  &=& -4 \sum_{A'B'}\frac{1}{(E^{(0)}_{A'}-E_{A}^{(0)}) + (E_{B'}^{(0)}-E^{(0)}_{B})}  \nonumber \\
&& \times \sum_{qq'} \left[\frac{\braket{ \Psi_A^{(0)} | \hat Q^{(1)}_{q} | \Psi_{A'}^{(0)}} \braket{ \Psi_B^{(0)} | \hat Q^{(1)}_{-q} | \Psi_{B'}^{(0)}}}{(1+q)!(1-q)!}\right. \nonumber \\
&& \left.\times \frac{\braket{ \Psi_{A'}^{(0)} | \hat Q^{(1)}_{-q'}| \Psi_{A}^{(0)}}\braket{ \Psi_{B'}^{(0)}| \hat Q^{(1)}_{q'}|\Psi_{B}^{(0)}}}{(1+q')!(1-q')!}\right],
\end{eqnarray}
where $E_{A}^{(0)}$ and $E_{B}^{(0)}$ are the molecular and atomic asymptotic energies at $R\rightarrow \infty$. All projections of the dipole tensors $\hat Q^{(1)}_{q} $ are taken into account.  Primed particle labels refer to intermediate states in the summation. Every intermediate rovibrational state  $\ket{\Psi_{A'}^{(0)}}$ in ground and excited electronic potentials is taken into account, as explained below. $M$ is the projection of the total angular momentum of the molecule along the internuclear axis. For alkali-metal atoms, we take into account all possible intermediate states $\ket{\Psi_{B'}^{(0)}}\equiv\ket{(n'^2L') {j'}m'}$ up to convergence of $C_6$. 
 
Following Ref. \cite{Spelsberg}, we rewrite the sum-over-states in Eq. (\ref{eq:2do-order}) in a more convenient form using the identities 

\label{eq:identities}
\begin{equation}
\frac{1}{a+b} = \frac{2}{\pi}\int_0^{\infty}d\omega\frac{ab}{(a^2+\omega^2)(b^2+\omega^2)},
\label{eq:upwards}
\end{equation}
and

\begin{equation}
\frac{1}{a-b} = \frac{2}{\pi}\int_0^{\infty}d\omega\frac{ab}{(a^2+\omega^2)(b^2+\omega^2)}+\frac{2a}{a^2-b^2},
\label{eq:downwards}
\end{equation}
which hold for positive real parameters $a$ and $b$. For molecular states, we set $ a=E_{\gamma' v'J'}-E_{X^1\Sigma vJ}$, where $\gamma'$ labels an intermediate electronic state. For atomic states, we set $b=E_{n'L'j'}-E_{nLj}$ for upward transitions ($E_{n'L'j'}>E_{nLj}$), and $b = E_{nLj}-E_{n'L'j'}$ for downward transitions ($E_{n'L'j'}<E_{nLj}$). Atom-molecule $C_6$ coefficients can thus be written as

\begin{eqnarray}\label{eq:C6}
C_6 &=& -\sum_{q, q'} K(q,q') \left\{ \int_0^{\omega_{\rm cut}} \frac{d\omega}{2\pi}  \;\alpha_{-q-q'}^{mm}(i\omega)\,\alpha_{qq'}^{MM}(i\omega) \right. \nonumber\\
&&+ \sum_{n''L''j''m''} \Theta(-\Delta E_{n''L''j''}) \alpha_{qq'}^{MM}(\Delta E_{n''L''j''})\nonumber\\
&&\;\;\;\;\;\;\times \left. \mathcal{T}_{nLjm}(n''L''j''m'') \right\},
\end{eqnarray}
where $K(q,q')\equiv 4/[(1+q)!(1-q)!(1+q')!(1-q')!]$ and $\omega_{\rm cut}$ is a cut-off frequency. 

The  arguments of integral in Eq. (\ref{eq:C6}) are the dynamic atomic polarizability component $\alpha_{-q-q'}^{mm}(z)$ and the dynamic molecular polarizability component $\alpha_{qq'}^{MM}(z)$, each evaluated at the imaginary frequency $z=i\omega$.  The second term in the square bracket represents downward transitions in the atom, with $\Delta E_{n''L''j''} =  E_{n''L''j''} - E_{nLj}$. The Heaviside function $\Theta(x)$ enforces the downward character of the transitions that contribute to this term. These terms are weighted by the  product of the atomic transition dipole integrals
\begin{eqnarray}\label{eq:dipoleintegrals}
\mathcal{T}_{nLjm}(n''L''j''m'') &\equiv& \matelement{(n^2L)jm}{Q^{(1)}_{-q}}{(n''^2L''){j''}m''} \nonumber \\
&&\times \matelement{(n''^2L''){j''}m''}{Q^{(1)}_{q'}}{(n^2L){j}m}, \nonumber\\
\end{eqnarray}
and the molecular polarizability function $\alpha_{qq'}^{MM}(\omega)$ evaluated at the frequency $\omega=\Delta E_{n''L''j''}/\hbar$. In this work, we evaluate both the integral and the downward contributions to $C_6$ up to convergence within a cut-off $\omega_{\rm cut}$, as described below in more details. 

\subsection{Polarizability of atomic Rydberg states}\label{atomic}

The dynamic atomic polarizability $\alpha_{-q-q}^{mm}(\omega)$ in Eq. (\ref{eq:C6}) can be written for a general atomic state  $\ket{k}$ as
\begin{equation}\label{eq:polariza}
\alpha^{kk}_{pp'}(\omega) = \sum_l \frac{\braket{k| \hat{d}_p^{\dagger} |l} \braket{l| \hat{d}_{p'} |k}}{E_l - E_k - \omega} + \frac{\braket{k| \hat{d}_{p'} |l} \braket{l| \hat{d}_p^{\dagger} |k}}{E_l - E_k + \omega},
\end{equation}
where $\omega$ is the frequency of the dipolar response,  $\hat{d}_p$ is the $p$-th component of the electric dipole operator in the spherical basis, $E_k$ is the zero-th order energy of state $\ket{k}$, and the state summation runs over all other atomic states $\ket{l}$ in the spectrum, with $l \neq k$. For alkali-metal atoms, the relevant atomic states and energies are obtained by numerically solving the radial Schr\"{o}dinger equation $\left[ -{\nabla^{2}}/{2} + V_L(r) \right] \Phi_B(r) = E_{nLj} \Phi_B(r)$ with a pseudo potential $V_L(r)$ that describes the interaction of core electrons with a single valence electron at distance $r$ from the core (origin), including spin-orbit coupling. The angular part of the atomic wavefunctions correspond to spherical harmonics $Y_{jm}(\theta,\phi)$.  We solve for the radial wavefunction $\Phi_B(r)$ as in Ref. \cite{Singer}, with atomic energies given by $E_{nLj} = -{hcR_{\infty}}/{(n-\delta_{nLj})^2}$, where $hcR_{\infty} = 1/2$ is the Rydberg constant (in atomic units). The fine structure quantum defects $\delta_{nLj}$ used in this work are given in Appendix \ref{sec:quantum defects}, in terms of the Rydberg-Ritz coefficients for $^{85}$Rb and $^{133}$Cs atoms.

We use the atomic energies and radial wavefunctions to construct the sum-over-states in Eq. (\ref{eq:polariza}), for a desired atomic state $\ket{(n^2L) {j}m}$. For convenience, the angular parts of the dipole integrals are evaluated using angular momentum algebra \cite{Zare}. The non-vanishing elements of the polarizability can thus be written as

\begin{widetext}
\begin{eqnarray}\label{eq:atompol}
\alpha^{mm}_{-q-q}(\omega) = \sum_{n''L''}\sum_{j''m''} (-1)^{2j''+2j-m''-m+q+1} \Bigg[ \frac{2(E_{n''L''j''}-E_{nLj})}{(E_{n''L''j''}-E_{nLj})^2-\omega^2}(2j+1)(2L+1)(2j''+1)(2L''+1) \nonumber\\
\times\threejm{L''}{0}{1}{0}{L}{0}^2\sixj{L}{j''}{j}{L''}{s}{1}^2\threejm{j''}{-m''}{1}{q}{j}{m}^2|\bra{n''L''}er\ket{nL}|^2 \Bigg],
\end{eqnarray}
\end{widetext}
where circular and curly brakets correspond to $3j$ and $6j$ symbols \cite{Zare}, respectively. We use Eq. (\ref{eq:atompol}) to compute the non-zero components of the polarizability tensor for atomic Rydberg  states $\ket{(n^2L) {j}m}$ with $n \geq 15$, ensuring the convergence of the sum over intermediate states for each imaginary frequency $i\omega$ that is relevant in the evaluation of the $C_6$ integral in Eq. (\ref{eq:C6}). 

\subsection{Polarizability of alkali-metal dimers}\label{molecular}

The dynamic molecular polarizability needed for the evaluation of the $C_6$ integral is also given by an expression as in Eq. (\ref{eq:polariza})), but for eigenstates $\ket{k}$ and energies $E_k$  describing electronic, vibrational and rotational state of alkali-metal dimers. For transition frequencies $(E_l-E_{k})/\hbar$ up to near infrared ($\sim 1$ THz), only states within the ground electronic potential need to be included explicitly in the summation. The contribution of transitions between rovibrational states in different electronic states is taken into account separately, as explained in what follows.

In the space-fixed frame, Eq. (\ref{eq:polariza}) can be given an explicit form by introducing molecular states $\ket{\gamma,v,JM}$ and energies $E_{\gamma vJ}$, where $\gamma$ labels the electronic state. Molecular polarizability components for a rovibrational state $\ket{X^1\Sigma,v,JM}$ in the ground electronic state can thus be written as \cite{Herrera}

\begin{eqnarray}\label{eq:SF alpha}
\alpha_{qq'}^{X^1\Sigma vJM}(\omega) &=& \sum_{\gamma' v'J'M'} \Bigg[\frac{2 (-1)^q (E_{\gamma'v'J'}-E_{X^1\Sigma vJ})}{(E_{\gamma'v'J'}-E_{X^1\Sigma vJ})^2-\omega^2}  \nonumber\\
&&\times \braket{X^1\Sigma vJM|\hat Q^{(1)}_{q}|\gamma'v'J'M'}\nonumber\\
&&\times\braket{\gamma'v'J'M'|\hat Q^{(1)}_{-q'}|X^1\Sigma vJM}\Bigg].
\end{eqnarray}
As anticipated above, for transition frequencies up to the mid-infrared, the dominant contributions to the molecular polarizability come from rovibrational transitions within the ground electronic state. Therefore, Eq. (\ref{eq:SF alpha}) can be rewritten as

\begin{equation}\label{eq:molPol}
\alpha^{X^1\Sigma vJM}_{qq'}(\omega)= \alpha_{qq'}^{\rm rv}(\omega) + \alpha_{qq'}^{\rm el}(\omega),
\end{equation}
where $\alpha_{qq'}^{\rm rv}$ and $\alpha_{qq'}^{\rm el}$ are the rovibrational and electronic polarizabilities, respectively given by

\begin{eqnarray}
\alpha_{qq'}^{\rm rv}(\omega) &=& \sum_{v'J'M'} \Bigg[\frac{2 (-1)^q (E_{X^1\Sigma v'J'}-E_{X^1\Sigma vJ})}{(E_{X^1\Sigma v'J'}-E_{X^1\Sigma vJ})^2-\omega^2}  \nonumber\\
&&\times\braket{X^1\Sigma vJM|\hat Q^{(1)}_{q}|X^1 \Sigma v'J'M'}\nonumber\\
&&\times \braket{X^1 \Sigma v'J'M'|\hat Q^{(1)}_{-q'}|X^1\Sigma vJM}\Bigg]
\label{eq:RotAlpha}
\end{eqnarray}
and
\begin{eqnarray}
\alpha_{qq'}^{\rm el}(\omega) &=& \sum_{\gamma' \neq X^1\Sigma}\sum_{v'J'M'} \Bigg[\frac{2 (-1)^q (E_{\gamma'v'J'}-E_{X^1\Sigma vJ})}{(E_{\gamma'v'J'}-E_{X^1\Sigma vJ})^2-\omega^2}  \nonumber\\
&&\times\braket{X^1\Sigma vJM|\hat Q^{(1)}_{q}|\gamma'v'J'M'}\nonumber\\
&& \times\braket{\gamma'v'J'M'|\hat Q^{(1)}_{-q'}|X^1\Sigma  vJM}\Bigg].
\label{eq:EleAlpha}
\end{eqnarray}

For evaluating the molecular dipole integrals in Eqs. (\ref{eq:RotAlpha}) and (\ref{eq:EleAlpha}), the space-fixed $q$-component of the dipole operator is written in terms of the body-fixed $p$-components through the unitary transformation $\hat Q^{(1)}_q = \sum_{p}\mathcal{D}^{*(1)}_{qp}\hat Q^{(1)}_{p}$, where $\mathcal{D}^{*(1)}_{qp}$ is an element of the Wigner rotation matrix \cite{Zare}. Transforming to the molecule-fixed frame is convenient since the electronic and vibrational eigenfunctions are  given in the body-fixed frame by most quantum chemistry packages.  The non-vanishing terms of the rovibrational polarizability ($q'=q$) can thus be written as
\begin{eqnarray}\label{eq:polRov}
\alpha_{qq}^{\rm rv}(\omega) &=& \sum_{v'J'M'} (2J+1)(2J'+1) \frac{2(E_{ v'J'}-E_{ vJ})}{(E_{ v'J'}-E_{ vJ})^2-\omega^2}\nonumber\\
&&\times \threejm{J'}{0}{1}{0}{J}{0}^2\threejm{J'}{-M'}{1}{-q}{J}{M}^2\nonumber\\
&&\times |\bra{vJ}\hat Q_0^{(1)}\ket{v'J'}|^2, 
\end{eqnarray}
where a redundant electronic state label has been omitted. The nuclear dipole integrals $\bra{vJ}\hat Q_0^{(1)}\ket{v'J'}$ can be evaluated directly once the rovibrational wavefunctions $\ket{v J}$ are known. These are obtained by solving the corresponding nuclear Schr\"{o}dinger equation (i.e., vibrations plus rotations) using a discrete variable representation (DVR) as in Ref.  \cite{Colbert}, with potential energy curves (PEC) and Durham expansions for the energies $E_{vJ}$ given in Ref. \cite{Vexiau} for the alkali-metal dimers used in this work.

For diatomic molecules, the electronic contribution to the polarizability in Eq. (\ref{eq:molPol}) is fully characterized in the body-fixed frame by the components $\alpha^{\rm el}_{00}(\omega)$ and $\alpha_{1,1}^{\rm el}(\omega)=\alpha_{-1,-1}^{\rm el}(\omega)$  \cite{Bonin}, which define the parallel polarizability $\alpha_{\parallel}(\omega)=\alpha_{00}^{\rm el}(\omega)$ and the perpendicular polarizability $\alpha_{\perp}(\omega)=-\alpha^{\rm el}_{\pm1,\pm 1}(\omega)$, with respect to its symmetry axis. For frequencies  up to the near infrared, the dynamic electronic polarizability of alkali-metal dimers do not deviate significantly from their static values $\alpha_\parallel(0)$ and $\alpha_\perp(0)$. Accurate static electronic polarizabilities for several alkali-metal dimers can obtained from Ref. \cite{deiglmayr:2008-alignment}. Explicitly, the space-fixed polarizability tensor for alkali-metal dimers in the $^1\Sigma$ state is given by
\begin{eqnarray}
\alpha^{\rm el}_{qq}(\omega) &=& \sum_{J'M'} (2J+1)(2J'+1) \threejm{J'}{-M'}{1}{-q}{J}{M}^2 \nonumber \\
&&\times \left[ \threejm{J'}{0}{1}{0}{J}{0}^2 \alpha_{\parallel}+2 \threejm{J'}{1}{1}{-1}{J}{0}^2\alpha_\perp \right]. \nonumber\\
\label{eq:elePol}
\end{eqnarray}
For the rovibrational ground state Eq. (\ref{eq:elePol}) reduces to its isotropic value $\alpha_{\rm iso}^{\rm el}=(\alpha_{\parallel}+2\alpha_{\perp})/3$ for all $q$-components. It was shown in Ref. \cite{Vexiau}, that for frequencies up to $\sim10^3 $ THz, the isotropic electronic molecular polarizability can be accurately approximated by 
\begin{equation}\label{eq:alpha effective}
\alpha_{\rm iso}^{\rm el}(\omega) = \frac{2\omega_{\Sigma}\,d^2_{\Sigma}}{\omega^2_{\Sigma}-\omega^2} + \frac{2\omega_{\Pi}\,d^2_{\Pi}}{\omega^2_{\Pi}-\omega^2},
\end{equation}
where the parameters $\omega_{\Sigma}$ and $d_{\Sigma}$ are the effective transition energy and dipole moment associated with the lowest $\Sigma\rightarrow \Sigma$ transition. The parameters $\omega_{\Pi}$ and $d_{\Pi}$ are associated with the lowest $\Sigma\rightarrow \Pi$ transition. For the alkali-metal dimers used in this work, we take the parameters listed in Ref. \cite{Vexiau} to estimate the electronic contribution to the molecular polarizability in over the frequencies of interest.  

Finally, we compute directly the downward transition terms that contribute to $C_6$ in Eq. (\ref{eq:C6}) by evaluating the total molecular polarizability in Eq. (\ref{eq:molPol}) at the relevant atomic transition frequencies, with an explicit evaluation of the atomic dipole integrals in Eq. (\ref{eq:dipoleintegrals}).

\begin{figure}[t]
\includegraphics[width=0.38\textwidth]{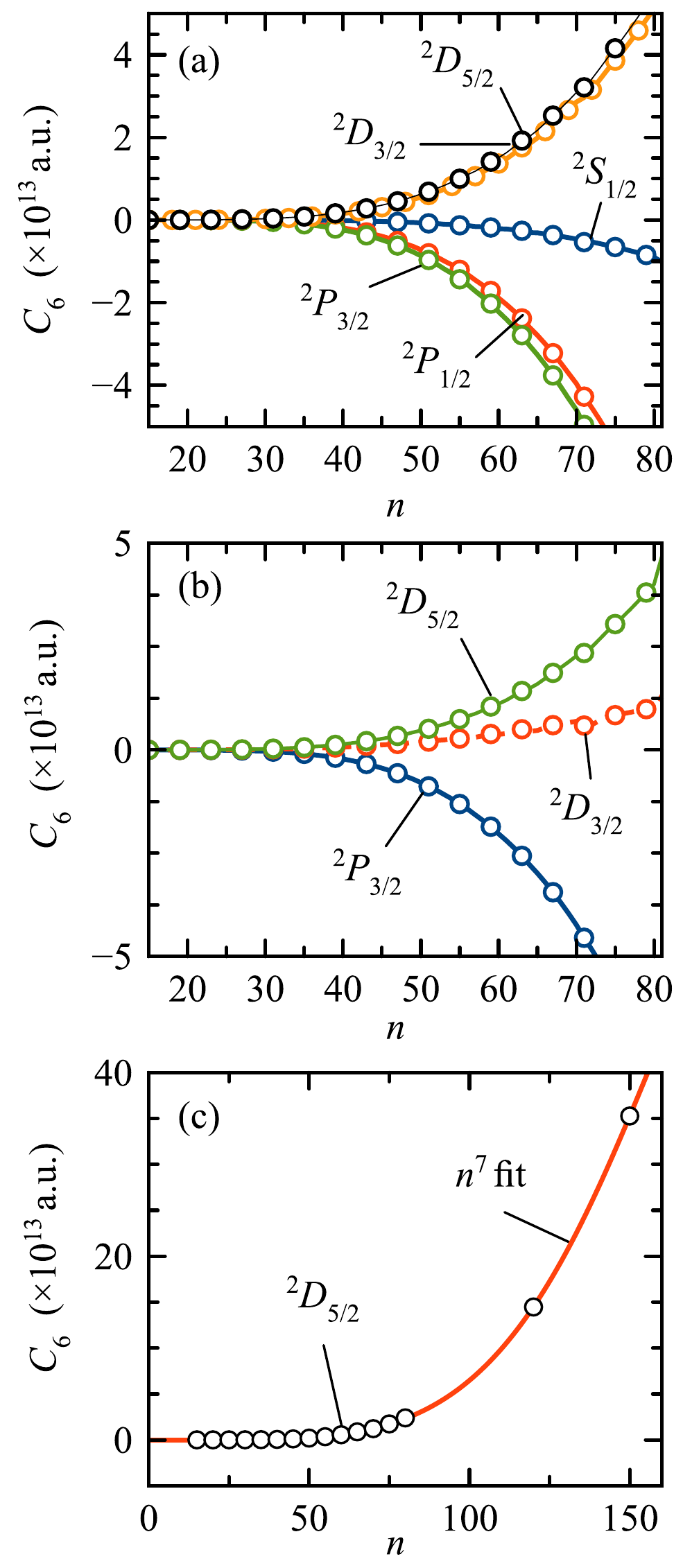}
\caption{$C_6$ dispersion coefficients as a function the atomic principal quantum number $n$ for the Cs-LiCs collision pair. Results are shown for several atomic Rydberg states $n^2L_j$. LiCs is in the electronic and rovibrational ground state. Panels show results for different total angular momentum projections  $\Omega = m + M$: (a) $|\Omega|=1/2$; (b) $|\Omega|=3/2$; (c) $|\Omega|=5/2$. Solid lines correspond to a fitted $n^7$ scaling.}
\label{fig:MolCs}
\end{figure}

\section{Results}\label{Results}

Motivated by recent atom-molecule co-trapping experiments \cite{McCarron2018,Segev2019}, we focus our analysis on the long-range interaction between two sets of collision partners: ({\it i}) $^{133}$Cs Rydberg atoms interacting with LiCs and RbCs molecules; ({\it ii}) $^{85}$Rb Rydberg atoms interacting with KRb, LiRb and RbCs molecules. We use Eq. (\ref{eq:C6}) to compute the $C_6$ coefficient of each atom-molecule pair considered, as a function of the principal quantum number $n$ of the atomic Rydberg state $n^{2}L_j$. We restrict our calculations to atomic states with $L\leq 2$. 

The total angular momentum projection  along the quantization axis 
\begin{equation}\label{eq:Omega}
\Omega= m+M,
\end{equation}
is a conserved quantity for an atom-molecule collision. For molecules in the rovibrational ground state ($J=0$), we thus have $\Omega=m$.  Below we present $C_6$ coefficients for each allowed value of $|\Omega|$. $C_6<0$ correspond to attractive interactions and $C_6>0$ describe repulsive potentials.

\subsection{Cesium + Molecule}

In Fig. \ref{fig:MolCs}, we plot the $C_6$ coefficients for $^{133}$Cs Rydberg states $n^{2}L_j$ interacting with  LiCs molecules in the rovibrational ground state, as a function of the atomic principal quantum number $n$, for all allowed values of $\Omega$. For concreteness, we restrict the atomic quantum numbers to the range $15\leq n\leq 150$, for $L\leq 2$.   

For Cs Rydberg atoms in $^2S_{1/2}$, $^2P_{1/2}$ and $^2P_{3/2}$ states, the interaction with LiCs molecules is attractive over the entire range of $n$ considered. As discussed in more detail below, this is due to the positive character of the atomic  and molecular polarizability functions at imaginary frequencies $\alpha(i\omega)$, which determine the value of the integral term in Eq.  (\ref{eq:C6}). On the other hand, Cs atoms in $^2D_{3/2}$ and $^2D_{5/2}$ Rydberg states give rise to repulsive $1/R^6$ potentials. This repulsive character of the atom-molecule interaction is due to the predominantly negative atomic polarizability function $\alpha(i\omega)$, while the molecular polarizability function remains positive. This is consistent with $n^2D$ Rydberg states having negative static polarizabilities $\alpha_{00}(\omega=0)$ \cite{Yerokhin}.
For both attractive and repulsive interactions, the magnitude of $C_6$ scales as $\sim n^7$ over a wide range of $n$, as shown explicitly in Fig. \ref{fig:MolCs}c. 

The $C_6$ coefficients for the Cs-RbCs collision pair exhibit the same qualitative behavior as the Cs-LiCs case, with repulsive potentials for $^2D_j$ states and attractive interaction for $^2S_j$ and $^2P_j$ Rydberg states.  We provide the complete list of all $C_6$ coefficients computed for the Cs-LiCs and Cs-RbCs collision partners  in the Supplementary Material.

\begin{figure}[t]
\includegraphics[width=0.40\textwidth]{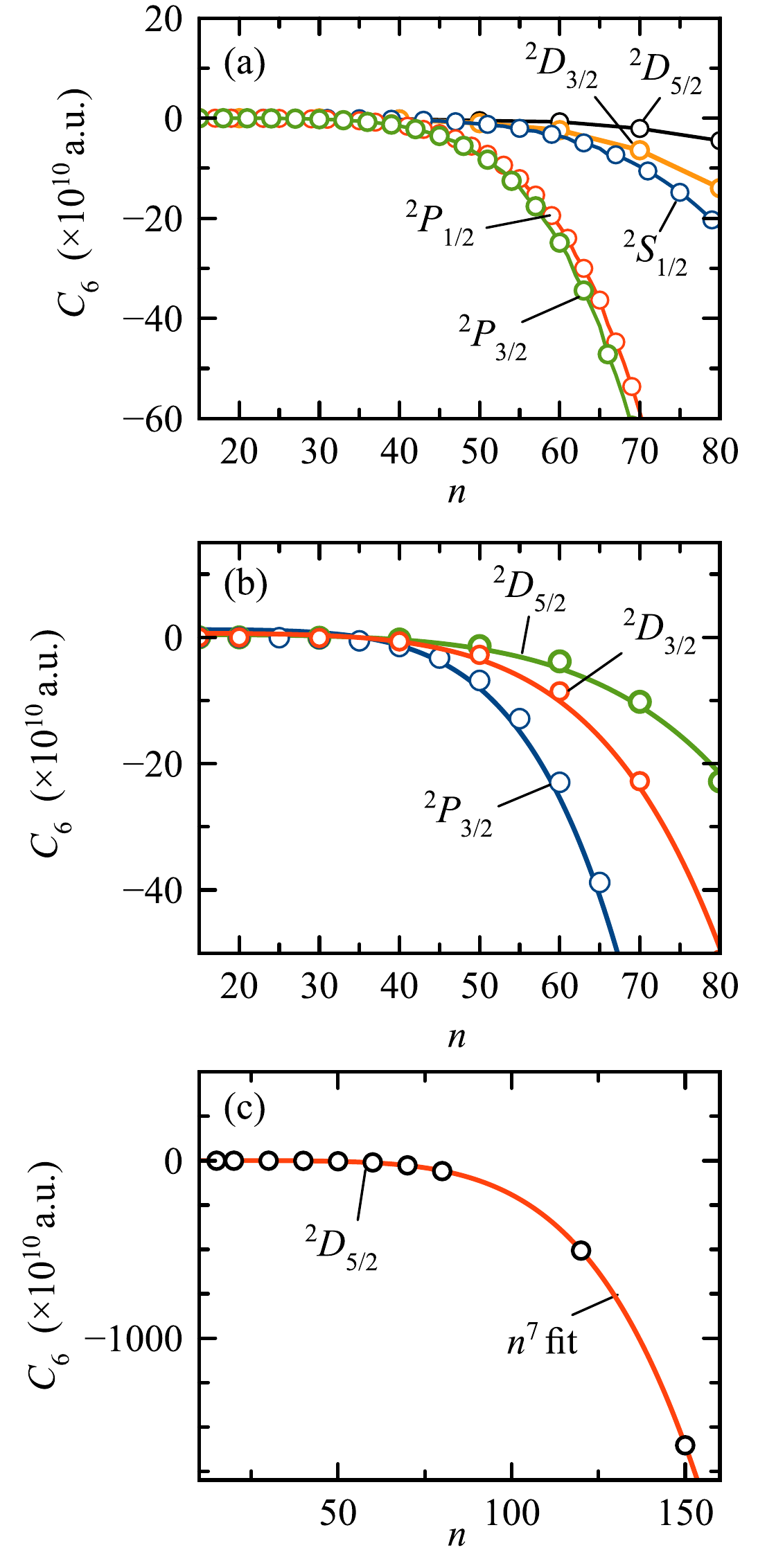}
\caption{$C_6$ dispersion coefficients as a function the atomic principal quantum number $n$ for the Rb-KRb collision pair. Results are shown for several atomic Rydberg states $n^2L_j$. KRb is in the electronic and rovibrational ground state. Panels show results for different total angular momentum projections  $\Omega = m + M$: (a) $|\Omega|=1/2$; (b) $|\Omega|=3/2$; (c) $|\Omega|=5/2$. Solid lines correspond to a fitted $n^7$ scaling.}
\label{fig:MolRb}
\end{figure}

\subsection{Rubidium + Molecule }

In Fig. \ref{fig:MolRb}, we plot the $C_6$ coefficients for $^{85}$Rb Rydberg states $n^2L_j$ interacting with  KRb molecules in the rovibrational ground state, as a function of the atomic principal quantum number $n$, for $L\leq 2$. The results resemble those of the Cs-LiCs pair with $^2S_{1/2}$, $^2P_{1/2}$ and $^2P_{3/2}$ atomic Rydberg states giving rise to attractive $1/R^6$ potentials that scale as $\sim n^7$, as show explicitly in Fig. \ref{fig:MolRb}b for $^2P_{3/2}$ states. In this case, $^2D_j$ states do not give repulsive potentials. 
 
The $C_6$ coefficients for the interaction of Rb Rydberg atoms with RbCs and LiRb molecules exhibit the same qualitative behavior as the Rb-KRb case, giving attractive interaction for $^2S_j$, $^2P_j$, and $^2D_j$ Rydberg states.  We provide the complete list of all $C_6$ coefficients computed for the Cs-LiCs and Cs-RbCs collision partners  in the Supplementary Material.

\subsection{Scaling of $C_6$ with $n$}

For all the atom-molecule pairs considered in this work, we fit the computed $C_6$ coefficients as a function of the atomic principal quantum number $n$ to the polynomial 
\begin{equation}\label{eq:fitting}
C_6 = \gamma_0+ \gamma_{4} \,  n^4 + \gamma_{6} \, n^6 + \gamma_{7} \, n^7.
\end{equation}
This scaling is valid in the range $n\approx 40-150$, with a fit quality that improves with increasing $n$. We list the fitting coefficients for Cs-LiCs and Cs-RbCs pairs in Table \ref{tab:Cs fit} for all atomic angular momentum states considered. The corresponding fitting coefficients for the collision pairs Rb-KRb, Rb-LiRb, and Rb-RbCs, are given in Table \ref{tab:Rb fit}. The  $n^7$ scaling found for $C_6$, is the same scaling of the static polarizability of Rydberg atoms \cite{Gallagher}. This suggests that the long-range interaction potential is dominated by the giant Rydberg polarizability, as expected.

\newcolumntype{L}{>{\raggedright\arraybackslash}X}%
\newcolumntype{R}{>{\raggedleft\arraybackslash}X}%
\newcolumntype{C}{>{\centering\arraybackslash}X}%

\begin{table*}[t]
\begin{tabularx}{0.9\textwidth}{| C | C | C | C | C C C C|}
\hline
 {\rm Molecule}& $L$ & $j$ & $|\Omega|$ & $\gamma_0$ & $\gamma_4$ & $\gamma_6$ & $\gamma_7$ \\
\hline
\multirow{9}{*}{LiCs} & $S$ & $1/2$ & $1/2$ & $1.518[11]$ & -$1.035[5]$ & -$30.94$ & $0.1630$ \\
                         \cline{2-8}
                      	 & \multirow{3}{*}{$P$} & $1/2$ & $1/2$ & $2.104[12]$ & -$1.984[6]$ & $162.1$ & -$1.606$ \\
                         \cline{3-8}                 	                                              	                        
                         &                 		& \multirow{2}{*}{$3/2$} & $1/2$ & $2.307[12]$ & -$2.366[6]$ & $238.2$ & -$2.375$ \\                                                                           
                         &                  	&						 & $3/2$ & $2.149[12]$ & -$2.186[6]$ & $225.5$ & -$2.236$ \\                                                                          
                         \cline{2-8}                         
						 & \multirow{5}{*}{$D$} & \multirow{2}{*}{$3/2$} & $1/2$ & -$1.428[12]$ & $1.708[6]$ & -$226.6$ & $1.879$ \\
                         &                 		&						 & $3/2$ & -$5.864[11]$ & $7.756[5]$ & -$134.4$ & $1.090$ \\
                                               	\cline{3-8}                                                   	                     	
                         &                  	& \multirow{3}{*}{$5/2$} & $1/2$ & -$1.431[12]$ & $1.909[6]$ & -$289.5$ & $2.469$ \\                                                                           
                         &                 		&						 & $3/2$ & -$1.081[12]$ & $1.462[6]$ & -$234.1$ & $1.973$ \\
                         &                 		&						 & $5/2$ & -$3.809[11]$ & $5.716[5]$ & -$123.2$ & $0.9800$ \\
\hline
\multirow{9}{*}{RbCs} & $S$ & $1/2$ & $1/2$ & -$2.907[10]$ & $2.757[4]$ & -$13.25$ & $0.05058$ \\
                         \cline{2-8}
                      	 & \multirow{3}{*}{$P$} & $1/2$ & $1/2$ & -$3.542[11]$ & -$1.737[5]$ & -$3.808$ & -$0.4723$ \\
                       	 \cline{3-8}
                         &                 		& \multirow{2}{*}{$3/2$} & $1/2$ & $4.866[11]$ & -$2.753[5]$ & $18.15$ & -$0.7826$ \\
                         &                  	&						 & $3/2$ & $4.425[11]$ & -$2.588[5]$ & $19.43$ & -$0.7408$ \\
                         \cline{2-8}
						 & \multirow{5}{*}{$D$} & \multirow{2}{*}{$3/2$} & $1/2$ & -$3.906[11]$ & $2.610[5]$ & -$31.40$ & $0.5847$ \\
                         &                 		&						 & $3/2$ & -$2.156[11]$ & $1.565[5]$ & -$26.86$ & $0.3579$ \\
                                               	\cline{3-8}                                               	       
                         &                  	& \multirow{3}{*}{$5/2$} & $1/2$ & -$4.818[11]$ & $3.501[5]$ & -$53.47$ & $0.8316$ \\
                         &                 		&						 & $3/2$ & -$3.846[11]$ & $2.821[5]$ & -$45.57$ & $0.6657$ \\
                         &                 		&						 & $5/2$ & -$1.901[11]$ & $1.458[5]$ & -$29.63$ & -$0.3334$ \\
\hline
\end{tabularx}
\caption{Parameters for the fitting $C_6 = \gamma_0+ \gamma_{4} \,  n^4 + \gamma_{6} \, n^6 + \gamma_{7} \, n^7$, for selected atom-molecule pairs involving $^{133}$Cs atoms in Rydberg states $n^2L_j$, interacting with LiCs and RbCs molecules in the ground electronic and rovibrational state. $\Omega=m$ is the total angular momentum projection of the collision pair. $C_6$ is in atomic units ($a_0^3$). The fitting is accurate in the range $n=40-150$. The notation $A[x]$ means $A \times 10^x$.}
\label{tab:Cs fit} 
\end{table*}

\newcolumntype{L}{>{\raggedright\arraybackslash}X}%
\newcolumntype{R}{>{\raggedleft\arraybackslash}X}%
\newcolumntype{C}{>{\centering\arraybackslash}X}%

\begin{table*}[t]
\begin{tabularx}{0.9\textwidth}{| C | C | C | C | C C C C|}
\hline
 {\rm Molecule} & $L$ & $j$ & $|\Omega|$ & $\gamma_0$ & $\gamma_4$ & $\gamma_6$ & $\gamma_7$ \\
\hline
\multirow{9}{*}{KRb} & $S$ & $1/2$ & $1/2$ & $4.340[9]$ & -$1905$ & -$0.5274$ & $4.695[$-$4]$ \\
                         \cline{2-8}
                      	 & \multirow{3}{*}{$P$} & $1/2$ & $1/2$ & $1.155[10]$ & $27.48$ & -$6.706$ & $0.02190$ \\
                       	 \cline{3-8}                      	 
                         &                 		& \multirow{2}{*}{$3/2$} & $1/2$ & $1.429[10]$ & -$840.9$ & -$7.419$ & $0.02381$  \\
                         &                  	&						 & $3/2$ & $1.227[10]$ & $9.555$ & -$7.089$ & $0.02286$ \\
                         \cline{2-8}
						 & \multirow{5}{*}{$D$} & \multirow{2}{*}{$3/2$} & $1/2$ & $4.907[9]$ & -$3284$ & $0.1596$ & -$1.883[$-$3]$ \\
                         &                 		&						 & $3/2$ & $6.531[9]$ & -$2307$ & -$2.039$ & $6.199[$-$3]$ \\
                                               	\cline{3-8}
                         &                  	& \multirow{3}{*}{$5/2$} & $1/2$ & $3.298[9]$ & -$2486$ & $0.4228$ & -$2.302[$-$3]$ \\                                                                           
                         &                 		&						 & $3/2$ & $4.481[9]$ & -$2337$ & -$0.5691$ & $1.173[$-$3]$ \\
                         &                 		&						 & $5/2$ & $6.782[9]$ & -$2026$ & -$2.521$ & $8.018[$-$3]$ \\
\hline
\multirow{9}{*}{LiRb} & $S$ & $1/2$ & $1/2$ & $1.381[11]$ & -$1.085[5]$ & $6.408$ & -$0.04418$ \\
                         \cline{2-8}
                      	 & \multirow{3}{*}{$P$} & $1/2$ & $1/2$ & $5.353[11]$ & -$4.224[5]$ & -$13.06$ & $0.01194$ \\
                       	 \cline{3-8}                      	 
                         &                 		& \multirow{2}{*}{$3/2$} & $1/2$ & $6.063[11]$ & -$4.839[5]$ & -$12.77$ & $1.718[$-$3]$ \\
                         &                  	&						 & $3/2$ & $5.598[11]$ & -$4.420[5]$ & -$14.57$ & $0.01386$ \\
                         \cline{2-8}
						 & \multirow{5}{*}{$D$} & \multirow{2}{*}{$3/2$} & $1/2$ & $1.091[11]$ & -$1.040[5]$ & $11.58$ & -$0.06377$ \\
                         &                 		&						 & $3/2$ & $2.100[11]$ & -$2.024[5]$ & $3.214$ & -$0.03349$ \\
                                               	\cline{3-8}
                         &                  	& \multirow{3}{*}{$5/2$} & $1/2$ & $8.086[10]$ & -$7.309[4]$ & $10.89$ & -$0.05510$ \\                                                                           
                         &                 		&						 & $3/2$ & $1.317[11]$ & -$1.237[5]$ & $7.670$ & -$0.04539$ \\
                         &                 		&						 & $5/2$ & $2.309[11]$ & -$2.225[5]$ & $1.254$ & -$0.02591$ \\
\hline
\multirow{8}{*}{RbCs} & $S$ & $1/2$ & $1/2$ & $1.316[10]$ & -$3897$ & -$2.743$ & -$7.242[$-$3]$ \\
                         \cline{2-8}
                      	 & \multirow{3}{*}{$P$} & $1/2$ & $1/2$ & -$1.576[10]$ & $5.049[4]$ & -$39.63$ & $0.07148$ \\
                       	 \cline{3-8}                      	 
                         &                 		& \multirow{2}{*}{$3/2$} & $1/2$ & -$1.225[10]$ & $5.381[4]$ & -$44.40$ & $0.07713$ \\
                         &                  	&						 & $3/2$ & -$1.640[10]$ & $5.347[4]$ & -$41.94$ & $0.07316$ \\
                         \cline{2-8}
						 & \multirow{4}{*}{$D$} & \multirow{2}{*}{$3/2$} & $1/2$ & $2.285[10]$ & -$1.490[4]$ & $1.265$ & -$0.01663$ \\
                         &                 		&						 & $3/2$ & $9.039[9]$ & $7469$ & -$12.55$ & $0.01839$ \\
                                               	\cline{3-8}
                         &                  	& \multirow{2}{*}{$5/2$} & $3/2$ & $5.808[10]$ & -$1.796[4]$ & -$1.099$ & -$0.01126$ \\
                         &                 		&						 & $5/2$ & $5.978[10]$ & -$2261$ & -$13.17$ & $0.01667$ \\
\hline
\end{tabularx}
\caption{Parameters for the fitting $C_6 = \gamma_0+ \gamma_{4} \,  n^4 + \gamma_{6} \, n^6 + \gamma_{7} \, n^7$, for selected atom-molecule pairs involving $^{85}$Rb atoms in Rydberg states $n^2L_j$, interacting with RbCs, LiRb and KRb molecules in the ground electronic and rovibrational state. $\Omega=m$ is the total angular momentum projection of the collision pair. $C_6$ is in atomic units ($a_0^3$). The fitting is accurate in the range $n=40-150$. The notation $A[x]$ means $A \times 10^x$.}
\label{tab:Rb fit} 
\end{table*}

\section{Discussion}\label{discussion}

\begin{figure}[t]
\includegraphics[width=0.38\textwidth]{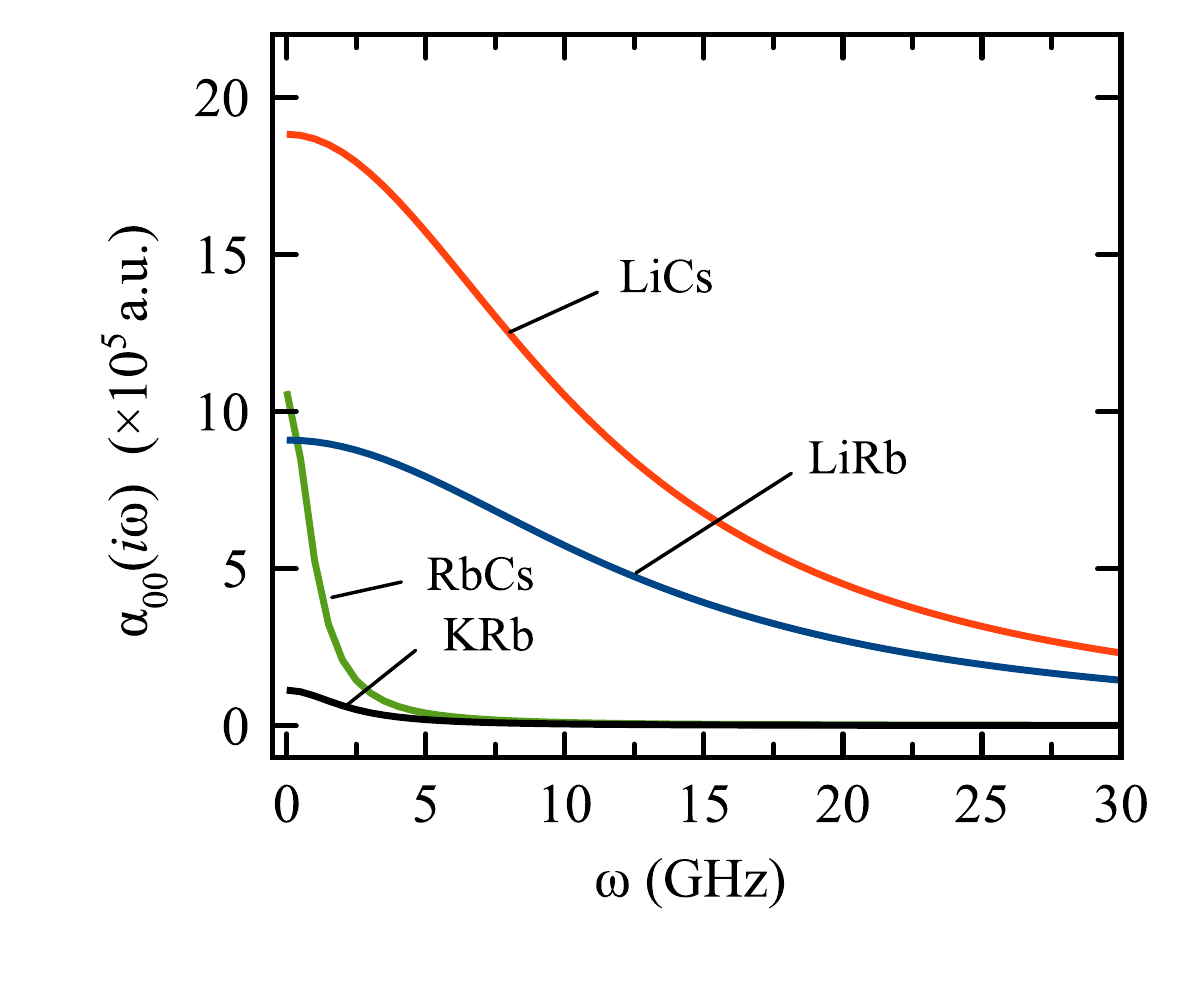}
\caption{Molecular polarizability function at imaginary frequency $\alpha_{00}(i\omega)$ for selected alkali-metal dimers in the electronic and rovibrational ground state $X^1\Sigma^+$, $\nu=J=M=0$.}  
\label{fig:MolPol} 
\end{figure}

\begin{figure}[t]
\includegraphics[width=.38\textwidth]{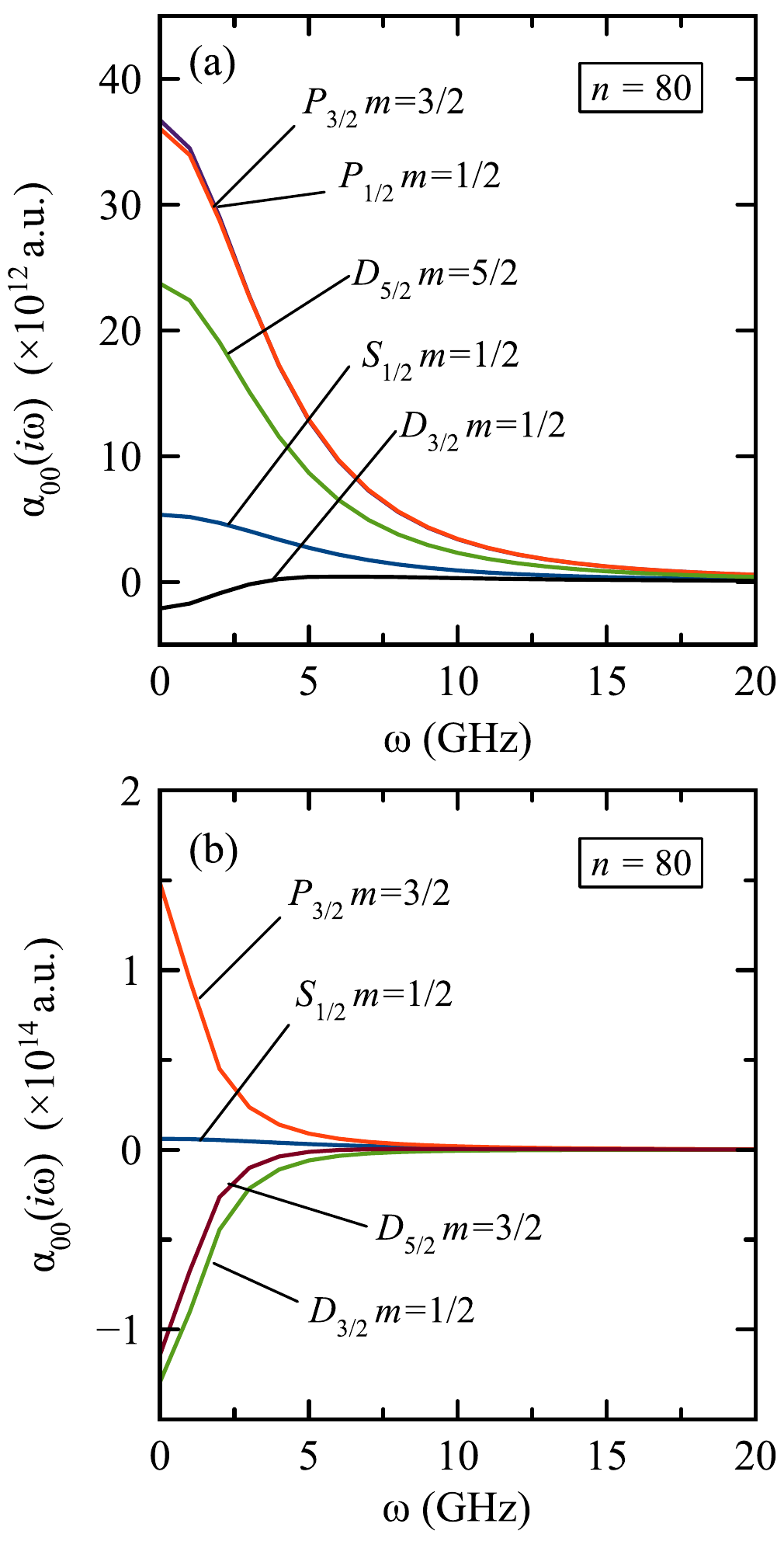}
\caption{Dynamic polarizability function at imaginary frequencies ($i\omega$) for $^{85}$Rb (a) and  $^{133}$Cs (b), for selected $n^2L_j$ states with $n=80$.}
\label{fig:AtoPol}
\end{figure}

Since the $C_6$ coefficients for the atom-molecule pairs listed in Tables \ref{tab:Cs fit} and \ref{tab:Rb fit} have yet to be experimentally measured, we  estimate their accuracy from other considerations. The first question to address is the importance of the contribution to $C_6$ of the downward transition terms in Eq. (\ref{eq:C6}).  We find that for all the atomic states $n^2L_j$  considered, the downward transition terms represent a negligible contribution to $C_6$ in comparison with the integral term that involves the atomic Rydberg polarizability function. 

This conclusion is valid provided we exclude resonant contributions to the downward transition term that involve evaluating the molecular polarizability at the atomic transition frequency $\Delta E_a=2B_e$, where $B_e$ is the rotational constant. High-$n$ Rydberg states with transition frequencies that are resonant with rotational excitation frequencies may instead contribute to energy exchange processes that scale as $1/R^3$, which can be avoided by careful selection of $n$ and $L$ quantum numbers.  After removing resonant contributions ($\Delta E_a=2B_e$) from the summation, the contribution of the integral term to $C_6$ in Eq. (\ref{eq:C6}) was found to be at least three orders of magnitude larger than the contribution of downward transition terms, for all the atom-molecule pairs studied in the range $n\geq 15$. One way to qualitatively understand this result is by comparing the $ n^7$ scaling of the static atomic polarizability $\alpha(0)$ versus the $ n^2$ scaling of the radial dipole integrals $\langle r^2\rangle^{1/2}$ for Rydberg states. The ratio between the integral (polarizability) and downward transitions (dipole) in Eq. (\ref{eq:C6}) can thus scale at least   as $n^3$, which gives a ratio of order $10^4$ for $n=50$ and $10^6$ for $n=100$.

\subsection{Error bounds on $C_6$ values}

After safely ignoring the atomic downward transition contributions to $C_6$ for $n>15$, we focus now on estimating the accuracy of the frequency integral contribution to Eq. (\ref{eq:C6}). The rovibrational structure and electrostatic response of most alkali-metal dimers in the ground $X^1\Sigma$ state is well-known from precision spectroscopy experiments and accurate {\it ab-initio} studies \cite{deiglmayr:2008-alignment,Deiglmayr:2008,Deiglmayr}. Therefore, the molecular polarizability function $\alpha_{qq}^{JM}$ in Eq. (\ref{eq:molPol}) is assumed to be known with very high precision in comparison with the atomic polarizability function \footnote{Our computed static molecular polarizabilities differ from the results in Ref. \cite{Vexiau} by less than $2\%$.}. In Fig. \ref{fig:MolPol} we plot the molecular polarizability function evaluated at imaginary frequencies $\alpha_{qq}^{JM}(i\omega)$ up to the microwave regime, for KRb, RbCs, LiRb and LiCs molecules. The figure shows the decreasing monotonic character of all molecular polarizability functions studied. As the frequency $\omega$ reaches the THz regime (not shown), all molecular functions $\alpha_{qq}^{JM}(i\omega)$ tend asymptotically to their isotropic static polarizabilities $\alpha_{\rm iso}^{\rm el}$ [Eq. (\ref{eq:alpha effective})], and remain constant over a large frequency range up to several hundred THz. In other words, over a broad frequency range up to $\sim 100$ THz, the contribution of the molecular polarizability to $C_6$ in Eq. \ref{eq:C6} is always positive, and can be considered to be bounded from above by its static value.

The accuracy of our computed atomic polarizability functions $\alpha_{qq}^{mm}(i\omega)$ is limited by the precision of the quantum defects used, which we take from spectroscopic measurements \cite{Singer}. 
  For the atomic Rydberg states considered, the polarizability functions obtained from Eq. (\ref{eq:atompol}) are predominantly monotonic as a function of frequency, although we found specific states $n^2L_j$ with non-monotonic frequency dependence. We illustrate this in Fig. \ref{fig:AtoPol}, where we show the polarizability functions $\alpha_{00}^{mm}(i\omega)$ for several $n^2L_j$ Rydberg levels of Rb and Cs atoms in the $n=80$ manifold. Panel \ref{fig:AtoPol}a shows that the projection $m=1/2$ of the $80^2D_{3/2}$ Rydberg state of Rb, the function $\alpha(i\omega)$ is negative in the static limit, then has a maximum at $\omega\approx 6.5$ GHz, from where it decays to a positive asymptotic value up to a cutoff frequency $\omega_{\rm cut}$ of a few  THz. This value at cutoff is five orders of magnitude smaller (not show) than the maximum in the microwave regime. In general, for all the atomic states considered, we find that  $|\alpha_{qq}^{mm}(i\omega)|$ is always bounded from above by its value at $\omega=0$. 

\begin{figure}[t]
\includegraphics[width=.42\textwidth]{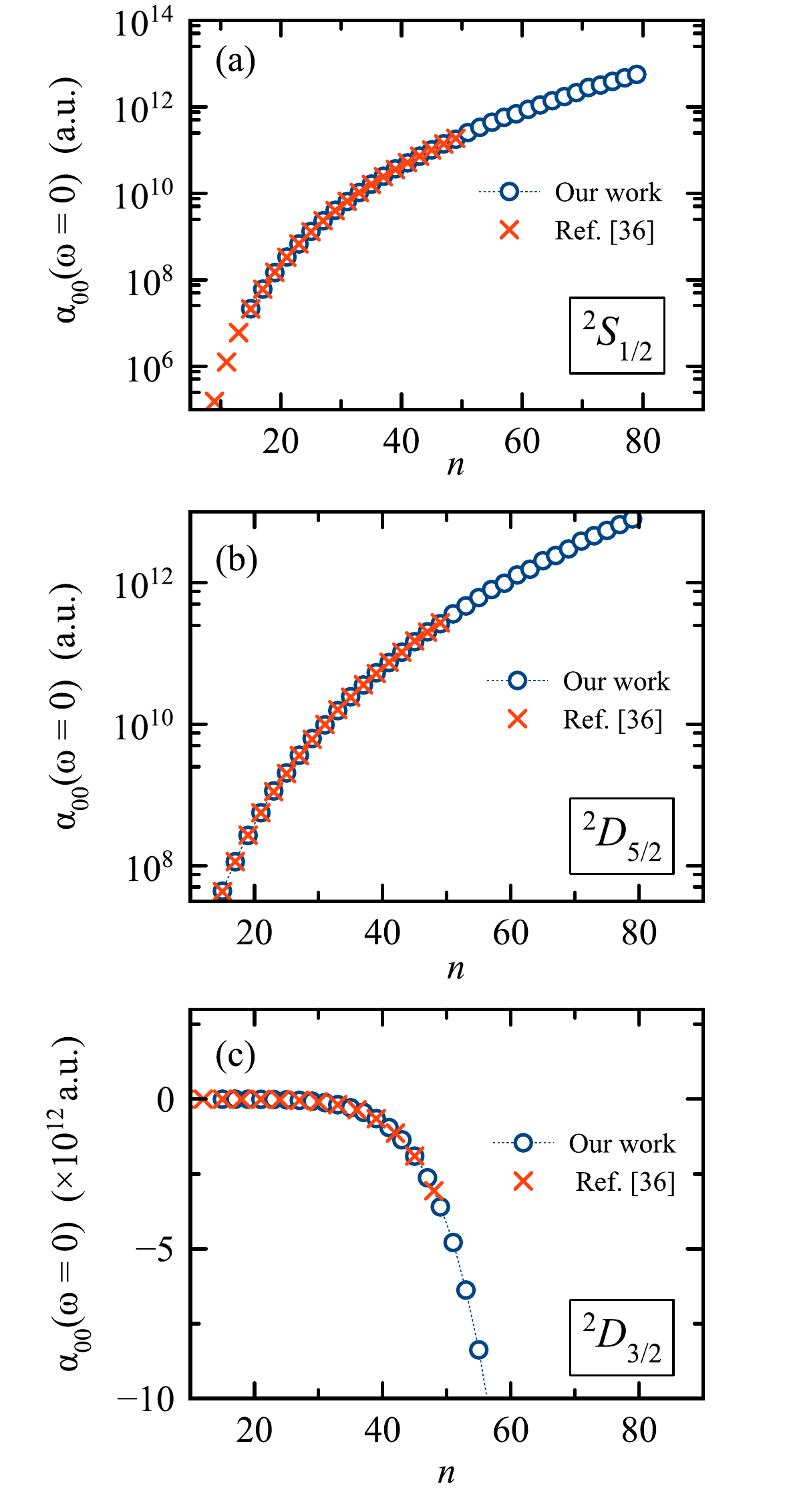}
\caption{Static polarizability of $^{133}$Cs atoms in selected angular momentum states (in units of $a_0^3$), as a function of the principal quantum number $n$, for { $^{133}$Cs} atoms in selected angular momentum states: (a) $\alpha_{00}$ component for $^2S_{1/2}$; (b) $\alpha_{00}$ component for $^2D_{5/2}$, $m=5/2$; (c) $\alpha_{00}$, $\alpha_{11}$, and $\alpha_{-1-1}$ components for $^2D_{3/2}$  $m=1/2$. Results for $\alpha_{00}$ from Ref. \cite{Yerokhin} are also shown.}
\label{fig:Cs pol} 
\end{figure}

The error of the computed $C_6$ coefficients can thus be estimated for $n\geq15$ as follows. Ignoring the downward transition terms, and the error in the molecular polarizability function, Eq. (\ref{eq:C6}) can be written as $\tilde C_6 =  C_6\pm\Delta C_6$, where $\tilde C_6$ is the dispersion coefficient obtained in our calculations, and the error is approximately given by
\begin{equation}\label{eq:C6 approx}
\Delta C_6 \approx-\sum_{q, q'} \frac{K(q,q')}{2\pi}  \int_0^{\omega_{\rm cut}} \frac{d\omega}{2\pi}  \;\Delta\alpha_{qq'}^{mm}(i\omega)\,\alpha_{-q-q'}^{MM}(i\omega),  
\end{equation}
where $\Delta \alpha_{qq'}^{mm}(i\omega)$ is the error in the atomic polarizability function evaluated at imaginary frequencies. We can assume that the order of magnitude of $C_6$ and $\Delta C_6$ is dominated by the $00$-components of the atomic and molecular polarizability functions. If we also assume that the relative error $\Delta \alpha_{qq}^{mm}(i\omega)/\alpha_{qq}^{mm}(i\omega)$ remains constant over all frequencies up to the cutoff $\omega_{\rm cut}$, and use the fact that $|\alpha(i\omega)|$ is bounded from above by its static value in the atomic and molecular cases, we can estimate an approximate error bound for $C_6$ as
\begin{equation}\label{eq:C6 error}
\left|\frac{\Delta C_6}{C_6}\right|\lesssim \left|\frac{\Delta \alpha_{00}^{mm}(0)}{\alpha_{00}^{mm}(0)}\right|.
\end{equation}
In other words, the accuracy of our $C_6$ calculations cannot expected to be better than the accuracy of the static atomic polarizability. The static polarizabilities of several Rydberg states of $^{85}$Rb and $^{133}$Cs are known from laser spectroscopy measurements in  static electric fields \cite{Zimmerman1979,Sullivan1986,Zhao2011}, and  also from precision calculations using state-of-the-art {\it ab-initio} pseudo-potentials \cite{Yerokhin}. Therefore, we can estimate $\Delta \alpha_{00}^{mm}(0)$ for several atomic Rydberg states $n^2L_j$, by comparing with available data. It proves convenient for comparisons to rewrite the atomic polarizability in Eq. (\ref{eq:atompol}) such that the Stark shift $\Delta E_{(nL)jm}$ of the Rydberg state $\ket{(n^2L)jm}$ in the presence of the electric field $E$ in the $Z$-direction can be written in the standard form \cite{Bonin}
\begin{equation}\label{eq:stark shift}
\Delta E_{(nL)jm}= -\frac{1}{2}\left[\alpha_0(j)+\alpha_2(j)\frac{3m^2-j(j+1)}{j(2j-1)}\right]E^2,
\end{equation}
 where $\alpha_0(j)$ is the scalar polarizability and $\alpha_2(j)$ the tensor polarizability. The factor in square brackets is equals to $\alpha_{00}^{mm}(0)$ in Eq. (\ref{eq:atompol}). 

In Fig. \ref{fig:Cs pol}, we plot the the static polarizability of $^{133}$Cs atoms in selected angular momentum states, as a function of the principal quantum number $n$.  As a standard, we use {\it ab-initio} results from Ref. \cite{Yerokhin}. Our computed values agree with the standard with very high accuracy. For example, the average relative errors over  the range $15\leq n\leq 50$ are $-0.02\%$ for $^2S_{1/2}$ states (panel \ref{fig:Cs pol}a), $+0.27\,\%$ for $^2D_{5/2}$ states with $m=5/2$ (panel \ref{fig:Cs pol}b), and $+0.13\,\%$ for $^2D_{3/2}$ states with $m=3/2$ (panel \ref{fig:Cs pol}c). For other Rydberg states of $^{133}$Cs we obtain similar accuracies.

\begin{figure}
\includegraphics[width=.42\textwidth]{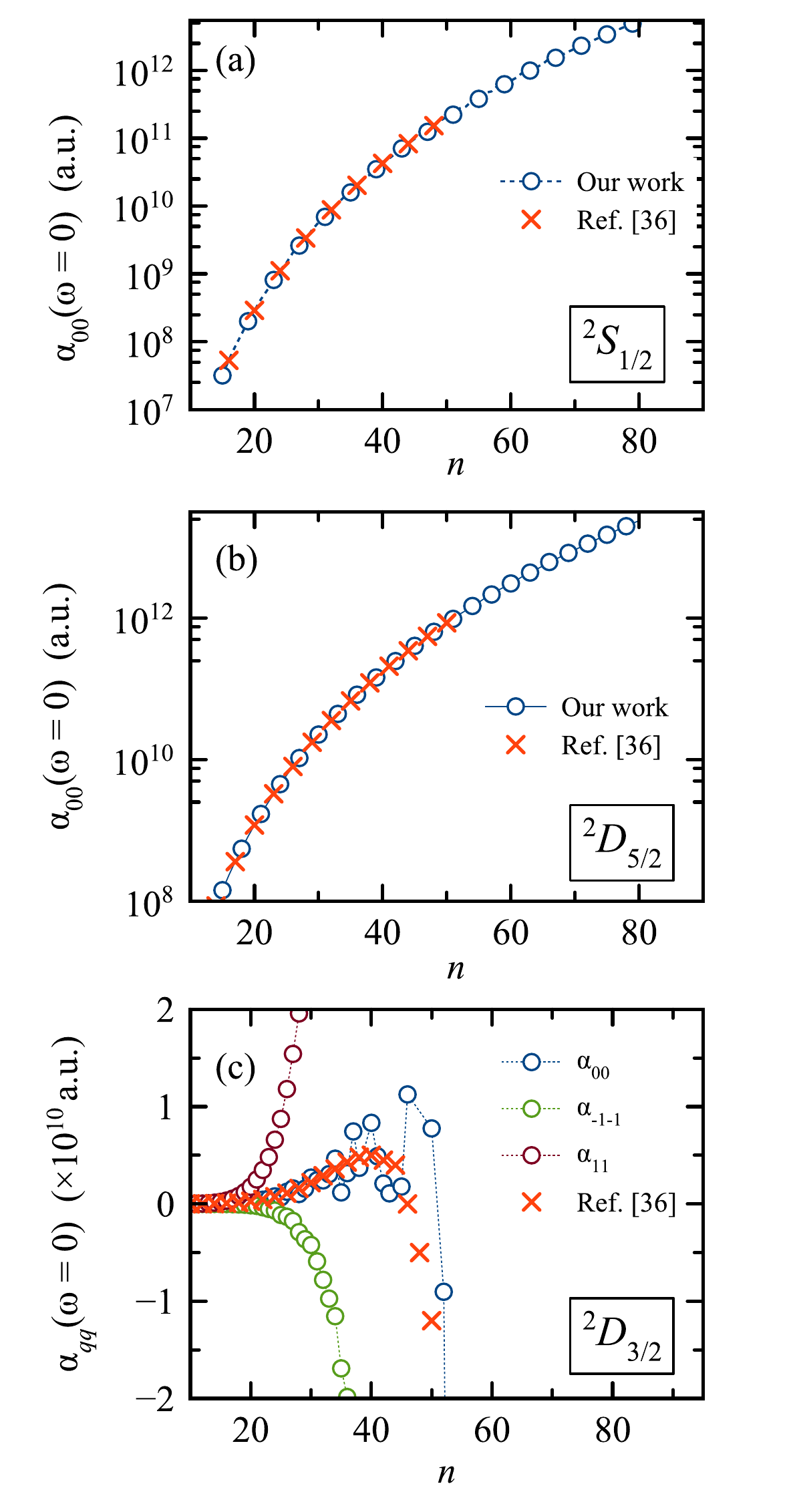}
\caption{Static polarizability of $^{85}$Rb atoms in selected angular momentum states (in units of $a_0^3$), as a function of the principal quantum number $n$: (a) $\alpha_{00}$ component for $^2S_{1/2}$; (b) $\alpha_{00}$ component for $^2D_{5/2}$, $m=5/2$; (c) $\alpha_{00}$, $\alpha_{11}$, and $\alpha_{-1-1}$ components for $^2D_{3/2}$  $m=1/2$. Results for $\alpha_{00}$ from Ref. \cite{Yerokhin} are also shown.}
\label{fig:Rb pol} 
\end{figure}

In Fig. \ref{fig:Rb pol}, we plot the the static polarizability of $^{85}$Rb atoms in selected angular momentum states, as a function of $n$. For $^{85}$Rb atoms, all the static polarizabilities we compute show excellent agreement with reference values (errors smaller than $1\%$), except for the Rydberg state $n^2D_{3/2}$ with $m=1/2$. For this atomic state, our results for $\alpha_{00}(0)$ have large relative errors around $n=45$, as Fig. \ref{fig:Rb pol}c shows. We can understand this by noting that for $j=3/2$ and $m=1/2$, Eq. (\ref{eq:stark shift}) reads $\Delta E= -(\alpha_0-\alpha_2)E^2/2$. For the $n^2D_{3/2}$ states of $^{85}$Rb, experiments show that $\alpha_0\approx \alpha_2$ over in the range $n=30-60$ \cite{Lai}, with $\alpha_0\lesssim \alpha_2$ in the higher end of this range. This is confirmed by the {\it ab-initio} results in Ref. \cite{Yerokhin}, which predict a change of sign in the static polarizability at $n=46$, from positive to negative. By separately comparing our results with experimental and theoretical values for $\alpha_0$ and $\alpha_2$ (not shown), we observe that our errors are of the same magnitude as the difference $\alpha_0-\alpha_2$ in the range $30 < n < 60$, which makes the atomic polarizability calculations unreliable for this particular tensor component ($\alpha_{00}$) and atomic quantum numbers. Errors can be traced to the empirical quantum defects used. We also show in Fig. \ref{fig:Rb pol}c that over the same range of $n$ in which $\alpha_{00}(0)$ exhibits large relative errors, other polarizability components that do not change sign behave smoothly.

Another possible source of error in our $C_6$ calculations is the choice of the high frequency cutoff $\omega_{\rm cut}$ in the numerical integration of Eq. (\ref{eq:C6}). For every atomic Rydberg states considered, we tested the numerical convergence of the integration by increasing the value of the cutoff until the relative change $\delta C_6/C_6$ was smaller than a predefined tolerance value $\varepsilon$. For atom-molecule pairs involving both  $^{85}$Rb and $^{133}$Cs atoms, the polarizability integral converges faster with increasing cutoff for intermediate and high values of $n\gtrsim 30$, in comparison with low-$n$ states. The latter result in slower integral convergence. We converged all our $n\approx 15$ integrals at $\omega_{\rm cut}=3$ THz with a tolerance $\varepsilon=0.01$, which ensures convergence over an entire range of $n$.

\subsection{Effect of the molecular dipole moment}

\begin{figure}[h]
\includegraphics[width=0.45\textwidth]{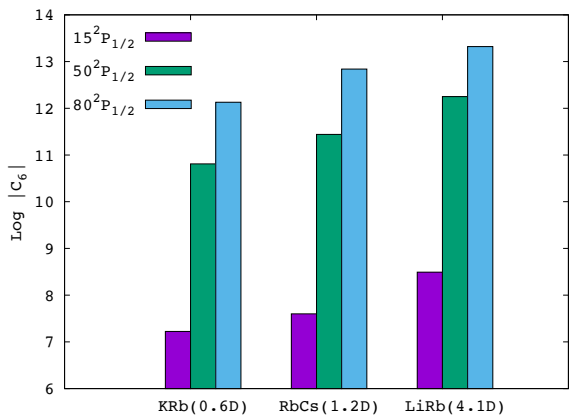}
\caption{Bar plots ${\rm log}|C_6|$ for $n=15,50,80$ for atom-molecule pairs involving $^{85}$Rb atoms in the $n^2P_{1/2}$ state with KRb, RbCs and LiRb molecules in the rovibrational ground state. The permanent dipole moment of each molecule is shown in parenthesis on the horizontal axis \cite{deiglmayr:2008-alignment}.}
\label{DipMol} 
\end{figure}

In Fig. \ref{DipMol} we show the increase in the magnitude of $C_6$ as the permanent dipole moment of alkali-metal dimers increases, for selected states $n^2P_{1/2}$ of $^{85}$Rb. The $C_6$ coefficient for the Rb-LiRb pair is larger than the corresponding values for RbCs and KRb, which have a smaller dipole moment. The same trend also holds for other $n^2L_{j}$ states of $^{85}$Rb, and for atom-molecule pairs involving $^{133}$Cs atoms.

\section{Conclusion}

The characteristic length scale for the van der Waals interaction between a Rydberg atom and a ground state alkali-metal dimer is the LeRoy radius $R_{LR}$ \cite{LeRoy}, given by the average root-mean-square electron radius of the collision pair. For alkali-metal atoms in low-lying Rydberg states ($n\approx 15-20$), the typical mean atomic radius can be on the order of $100-1000 \,a_0$, where $a_0$ is the Bohr radius, thus exceeding by orders the typical size of the electron radius of ground state molecules of only a few Bohr radii. The ratio between atomic and molecular radial distances further increases with $n$. For the atomic states considered in this work, the van der Waals length is thus dominated by the LeRoy radius of the Rydberg atom $R_{nL}\equiv {\langle n^2L|r^2 |n^2L\rangle}^{1/2}$. Given the $n^2$ scaling of the Rydberg radius and the $n^7$ scaling of the atom-molecule $C_6$ coefficients, the van der Waals energy should thus approximate scale as $U_{\rm vdW}\equiv C_6/R_{nL}^{6}\sim n^{-5}$. We find this scaling to be most accurate for $n\gtrsim 50$. 

From the values of $C_6$ listed in Tables \ref{tab:Cs fit}--\ref{tab:Rb fit}, the van der Waals energy $U_{\rm vdW}$ can be estimated in absolute units. For example, for the LiCs--Cs system with $^{133}$Cs in the $n^2D_{5/2}$ state and $\Omega=5/2$, the van der Waals potential is repulsive (Fig. \ref{fig:MolCs}c), with a  collisional barrier reaching $U_{\rm vdW}\approx 29$ MHz for $n=20$. This should be sufficient to avoid short-range collisions for  atom-molecule pairs with relative kinetic energy up to 150 $\mu$K. By increasing the atomic quantum number to $n=40$, the potential barrier drops to $U_{\rm vdW}\approx 0.41$ MHz for the same collision pair. Our results thus suggest that given a specific atom-molecule system of experimental interest, it is possible to find an atomic Rydberg state that gives an attractive or repulsive potential with a desired interaction strength. 

We can extend the formalism in this work to also obtain van der Waals coefficients for excited rovibrational states of alkali-metal dimers. In this case, $C_5$ coefficients do not vanish in general \cite{Lepers}. The interplay between $C_5$ and $C_6$ with opposite signs at long distances can possibly lead to long-range potential wells that can support Rydberg-like metastable bound states accessible in photoassociation spectroscopy \cite{Lepers3,Lepers2,Lepers,Perez2015}.

Repulsive van der Waals interactions may be used for sympathetic cooling of alkali-metal dimers via elastic collisions with ultracold Rydberg atoms. Since inelastic and reactive ultracold collisions \cite{Julienne, Idziaszek} can lead to spontaneously emitted photons carrying energy away from a trapped system \cite{Zhao}, it should be possible to measure the elastic-to-inelastic scattering rates and follow the thermalization process of a co-trapped atom-molecule mixture. Attractive van del Waals potentials can be exploited to form long-range alkali-metal trimers via photoassociation \cite{Shaffer2018}.

\acknowledgements
VO thanks support by Conicyt through grant Fondecyt Regular No. 1181743. FH is supported by Conicyt through grant REDES ETAPA INICIAL, Convocatoria 2017 REDI170423, Fondecyt Regular No. 1181743, and by  Iniciativa Cient\'{i}fica Milenio (ICM) through the Millennium Institute for Research in Optics (MIRO).

\appendix

\section{Quantum defects}\label{sec:quantum defects}

The quantum defects used in this work are collected from Ref. \cite{Singer} in terms of the expansion 
\begin{equation}\label{eq:qdefects}
\delta_{nLj}=a+\frac{b}{(n-a)^2}+\frac{c}{(n-a)^2}+\frac{d}{(n-a)^2}+\frac{e}{(n-a)^2},
\end{equation}
where the Rydberg-Ritz coefficients $(a,b,c,d)$ are given in Table \ref{Rb} for $^{85}$Rb and Table \ref{Cs} for $^{133}$Cs atoms, together with the minimum value of $n$ for which the expansion is estimated to be valid. 

\setlength{\tabcolsep}{2pt} 
\renewcommand{\arraystretch}{1.3} 
\begin{table}[t]
\begin{tabular}{c|c c c c c}
\hline

$L_{j}$ & $a$ & $b$ & $c$ & $d$ & $n_{\rm min}$ \\ \hline

 $S_{1/2}$ & $3.1311804(10)$ & $0.1784(6)$ & -$1.8$ & $-$ & $14$ \\
 $P_{1/2}$ & $2.6548849(10)$ & $0.2900(6)$ & -$7.9040$ & $116.4373$ &  $11$ \\
 $P_{3/2}$ & $2.6416737(10)$ & $0.2950(7)$ & -$0.97495$ & $14.6001$ &  $13$ \\
 $D_{3/2}$ & $1.34809171(40)$ & -$0.60286(26)$ & -$1.50517$ & -$2.4206$ & $4$ \\
 $D_{5/2}$ & $1.34646572(30)$ & -$0.59600(18)$ & -$1.50517$ & -$2.4206$ & $4$ \\
 $F_{j}$ & $0.016312$ & -$0.064007$ & -$0.36005$ & $3.2390$ & $4$ \\ \hline

\end{tabular}
\caption{Rydberg-Ritz coefficients for $^{85}$Rb.}   
\label{Rb} 
\end{table}

\setlength{\tabcolsep}{2pt} 
\renewcommand{\arraystretch}{1.3} 
\begin{table}[t]
\begin{tabular}{c|c c c c c}
\hline

$L_{j}$ & $a$ & $b$ & $c$ & $d$ & $n_{\rm min}$ \\ \hline

 $S_{1/2}$ & $4.049352(38)$ & $0.238(7)$ & $0.24044$ & $0.12177$ &  $6$ \\
 $P_{1/2}$ & $3.5916(5)$ & $0.36(1)$ & $0.34284$ & $1.23986$ & $6$ \\
 $P_{3/2}$ & $3.5590(7)$ & $0.38(1)$ & $0.28013$ & $1.57631$ & $6$ \\
 $D_{3/2}$ & $2.475454(20)$ & $0.010(4)$ & -$0.43324$ & -$0.96555$ & $5$ \\
 $D_{5/2}$ & $2.466308(30)$ & $0.015(6)$ & -$0.43674$ & -$0.74442$ & $5$\\
 $F_{5/2}$ & $0.033587$ & -$0.213732$ & $0.70025$ & -$3.66216$ & $4$ \\ \hline

\end{tabular}
\caption{Rydberg-Ritz coefficients for $^{133}$Cs.}   
\label{Cs} 
\end{table}


\clearpage
\bibliography{rydberg}{}

\bibliographystyle{unsrt}

\end{document}